\documentstyle[epsf,amsmath,amssymb,amsbsy,amsthm]{elsart}
\numberwithin{equation}{section}
\theoremstyle{plain}
\newtheorem*{cnj}{Conjecture}
\theoremstyle{remark}
\newtheorem{rmk}{Remark}
 
\begin{document}
\begin{frontmatter}
  \title{Matched asymptotic solutions for the steady banded flow of the 
    diffusive Johnson-Segalman model in various geometries\thanksref{jnnfm}} 
  \thanks[jnnfm]{Submitted to \emph{Journal of Non-Newtonian Fluid Mechanics}}
  \author{O.  Radulescu\thanksref{or}}
  \and \author{P.~D.  Olmsted\thanksref{pdo}}
  \thanks[or]{\texttt{phyor@irc@leeds.ac.uk}}
  \thanks[pdo]{\texttt{p.d.olmsted@leeds.ac.uk}}
  \address{Department of Physics and
    Astronomy, and IRC in Polymer Science and Technology, University of
    Leeds, Leeds LS2 9JT, United Kingdom}
  \begin{abstract}
    We present analytic solutions for steady flow of the
    Johnson-Segalman (JS) model with a diffusion term in various
    geometries and under controlled strain rate conditions, using
    matched asymptotic expansions.  The diffusion term represents a
    singular perturbation that lifts the continuous degeneracy of
    stable, banded, steady states present in the absence of diffusion.
    We show that the stable steady flow solutions in Poiseuille and
    cylindrical Couette geometries always have two bands.  For Couette
    flow and small curvature, two different banded solutions are
    possible, differing by the spatial sequence of the two bands.
  \end{abstract}
\end{frontmatter}
\section{Introduction}\label{sec:intro}
Many experimental results confirm the possibility of shear banding,
i.e. the separation of bands of different shear rates and apparent
viscosities in the flow of various systems of wormlike micelles
(aqueous solutions of surfactants and salt
\cite{hoffmann81,berret94b,callaghan96,grand97,boltenhagen97b} or
organic solvent solutions of metallic complexes \cite{terech98}), or
lyotropic liquid crystals \cite{Bonn+98}.  This type of behaviour can
be explained as being the result of a constitutive instability,
suitably described by non-monotonic flow curves such as those arising
from the Johnson-Segalman (JS) \cite{johnson77} or Doi-Edwards models
\cite{doiedwards}. It is possible that similar constitutive
instabilities are responsible for spurt and extrudate distortions of
polymer melts \cite{mcleish86}.

The steady banded flow solutions of the JS model were previously
studied in planar \cite{renardy,espanol96}, Poiseuille
\cite{malkus90,malkus91,nohel90,geovla98,fyrilas99}, and cylindrical
Couette geometries \cite{greco97}. In all these geometries banded flow
solutions have continuous degeneracy
\cite{geovla98,fyrilas99,olmradlu99}, resulting from the indeterminacy
of the positions of the interfaces and of the number of bands.  The
stress value can be kinetically selected
\cite{mcleish87,espanol96,greco97,geovla98,fyrilas99}, and it may
differ from the top jump prediction of Refs.~\cite{malkus90,malkus91},
although the selected value depends on the flow history and on the
imposed shear rate \cite{olmradlu99}.  This is in contrast to
experiments on wormlike micelles that show a well defined stress
plateau in the flow curves: the total stress is history independent
and do not change with the shear rate. Furthermore, the evolution of
the stress during transient flow may be rather long, suggesting the
slow migration of the interface between bands to an equilibrium
position \cite{berret94b}.

Recently \cite{olmradlu99,jsplanar,eurorheo} we showed that, by
supplementing the JS model with a stress diffusion term, one can lift
the degeneracy of the steady flow and also account for the slow
migration of the interface. The origin of this diffusion term can be
justified in terms of the Brownian movement of polymer chains in an
inhomogeneous stress field and can be deduced from the Fokker-Planck
equation for a system of dumbbells \cite{elkareh89}. There is no
calculation, at present, for such a term in the wormlike micelle
system.  Similar results have been obtained by
Refs.~\cite{brunovsky94,spenley96} by introducing phenomenological
diffusion terms in toy constitutive models.  Although non-local terms
are not important in homogeneous flow, in banded flow a realistic
description of interfaces as strongly inhomogeneous regions requires
such terms.  This necessity has been noted by many authors
\cite{elkareh89,pearson94,olmsted92,olmstedlu97}.

Recent experiments \cite{MairCall97} on wormlike micelles in pipe
(Poiseuille) and cylindrical Couette flow geometries, or in cone and
plate \cite{BritCall97} using NMR microscopy found shear rate profiles
with two or three bands. Birefringence measurements in Couette
geometry \cite{DCC97,decruppe98} suggest the existence of only two
bands, although the relation between birefringence and shear rate is
hard to establish because it depends on microstructure.  Most
experiments on shear banding are performed in geometries imposing
inhomogeneous total stress (Couette, pipe flow, cone and plate).
Typically a stress quantity (\emph{e.g.} torque, or pressure drop, or
shear stress) is measured as a function of a rate of flow quantity
(\emph{e.g.} mean strain rate, flux through a pipe, wall speed), and
an engineering ``flow curve'' is constructed.  Under inhomogeneous
flow conditions the flow curves are not simply the local constitutive
relation of the fluid.  It is therefore important to assess
theoretically the question of how many bands can occur and which are
the flow curves for a given constitutive model and in various
geometries.

We present elsewhere numerical computations of flow curves in Couette
flow \cite{olmradlu99} for the JS-d model.  Here we show how matched
asymptotics techniques can be used to obtain analytic results on the
number of bands, interface width, and flow curves in various
geometries. The application of these techniques to the JS-d model
represents only an example, and the same method can be easily adapted
to other constitutive models.

The structure of the paper is the following. In Section
\ref{sec:model} we introduce the flow equations in various geometries.
The matched asymptotic solutions of these equations are presented in
Section \ref{sect:matched}.  In Section \ref{sect:uniqueness} we
discuss how the inner layer solution controls the history independence
and stability of the flow.  We apply our results in Section
\ref{sect:applications} to predict the shapes of flow curves and the
interface width. In Section \ref{sect:conclusions} we discuss the
results and suggest possible extensions.
\section{Flow equations in various geometries}\label{sec:model}
\subsection{General equations}
The dynamics of the diffusive Johnson-Segalman fluid is described by
two equations. The first is the momentum balance:
\begin{equation}
  \label{eq:NS}
  \rho\left(\partial_t + {\bf v}\!\cdot\!\boldsymbol{\nabla}\right){\bf v}
  = \boldsymbol{\nabla}\!\cdot\!{\bf T},
\end{equation}
where $\rho$ is the fluid density, ${\bf v}$ is the velocity field,
and ${\bf T} = -p\,{\bf I} + 2\eta{\bf D} + \boldsymbol{\Sigma}$ is
the total stress tensor where the pressure $p$ is determined by
incompressibility, $\Sigma$ is the viscoelastic stress carried by the
polymer strands, $\eta$ is the ``solvent viscosity'' and ${\bf D}$ is the
symmetric part of the velocity gradient. The second equation is a
constitutive relation for the polymer stress $\boldsymbol{\Sigma}$,
which we take to be the the JS constitutive model \cite{johnson77}
with an added diffusion term \cite{jsplanar,olmradlu99}:
\begin{equation}
  \label{eq:JS}
  \overset{\blacklozenge}{\boldsymbol{\Sigma}}  = {\cal
    D}\nabla^2\boldsymbol{\Sigma} + 2\frac{\mu}{\tau}{\bf D} - 
  \frac{1}{\tau}\boldsymbol{\Sigma},
\end{equation}
where $\mu$ is the ``polymer'' viscosity, $\tau$ is the relaxation
time, and ${\cal D\/}$ is the diffusion coefficient.

The time evolution of $\boldsymbol{\Sigma}$ is governed by the
Gordon-Schowalter (GS) time derivative \cite{gordon72},
\begin{equation}
\overset{\blacklozenge}{\boldsymbol{\Sigma}} = \left(\partial_t +
  {\bf v}\!\cdot\!\boldsymbol{\nabla}\right)\boldsymbol{\Sigma} -
\left(\boldsymbol{\Omega\Sigma} - \boldsymbol{\Sigma\,\Omega}\right) -
a \left({\bf D}\boldsymbol{\Sigma} +
  \boldsymbol{\Sigma}{\bf D}\right)
\end{equation}
where $a$ is the ``slip parameter'' and $\boldsymbol{\Omega}$ is the
antisymmetric part of the velocity gradient. For $a=1$ there is no
slip and the GS derivative becomes the upper convective derivative.
For $a=0$ (total slip) the fluid stress is not transmitted to the
polymer strands and these can only be oriented by the flow: the GS
derivative becomes the corotational derivative. In this work we
consider $|a| < 1$.

We study the case of a fixed average shear rate, which represents a
global constraint on the velocity gradient.  We discuss several shear
geometries: planar shear, slit and pipe Poiseuille flow, and
cylindrical Couette flow between concentric cylinders.  We consider
parallel stream lines in all geometries and leave the possibility of
secondary flow for further investigation.
\subsection{Planar  shear and Poiseuille flow}
Parallel stream line flow corresponds to ${\bf v}=v(y) {\bf e}_x$, for
planar shear between plates at positions $y=0,L$ and for Poiseuille
flow through a rectangular slit of infinite length and walls at
positions $y=\pm L$.

We introduce dimensionless variables, $ \hat V = - V
\frac{\tau}{L}\sqrt{1-a^2} \hat W = (\frac{1-a}{2} \Sigma_{yy} -
\frac{1+a}{2}\Sigma_{xx}) \frac{\tau}{\mu} , \hat{\dot{\gamma}} =
\dot{\gamma} \tau \sqrt{1-a^2}, \hat Z = (\frac{1-a}{2} \Sigma_{yy} +
\frac{1+a}{2}\Sigma_{xx}) \frac{\tau}{\mu} , \hat S = \Sigma_{yx}
\frac{\tau}{\mu}\sqrt{1-a^2}, {\hat D\/} = {\cal D\/}
\frac{\tau}{L^2}, f = \frac{\partial p}{\partial
  x}\frac{L\sqrt{1-a^2}}{\mu \tau}, \hat \sigma = T_{yx}
\frac{\tau}{\mu}\sqrt{1-a^2}, \hat t = t / \tau, \hat r = y / L. $
$V$ is the velocity at $y=0$ (the maximum absolute value), having
opposite sign to the shear rate and stress. With these definitions
$\hat V$ has the same sign as the reduced shear rate and stress.
In terms of these quantities the momentum balance reads:
\begin{equation}
\alpha \partial_{\hat t} \hat V = 
\partial_{\hat r}\hat S + \epsilon \partial^2_{\hat r}\hat V - f.
\label{eq:mom_bal}
\end{equation}
The dimensionless pressure gradient $f$ vanishes in planar shear,
$\epsilon = \frac{\eta}{\mu}$ is the retardation parameter (viscosity
ratio) and $\alpha= \frac{\rho L^2}{\mu \tau}$ is the ratio of the
Reynolds and the Weissenberg numbers. In shear-banding experiments
(\emph{e.g.} for wormlike micelles) $\alpha \sim 10^{-4} - 10^{-3}$,
therefore in this case one can neglect inertia and replace
Eq.~(\ref{eq:mom_bal}) by its creeping flow limit:
\begin{equation}
\partial_{\hat{r}} ( \epsilon \hat{\dot{\gamma}} + \hat S ) = 0
\label{eq:creep_plan}
\end{equation}
The solutions of Eq.~(\ref{eq:creep_plan}) are:
\begin{subequations}
\label{eq:stressplan}
  \begin{align}
    \hat \sigma(\hat r) &=const\qquad&\text{(planar shear)}
    \label{eq:stressplanar}\\[7truept]
    \hat \sigma(\hat r) &= f \hat r&\text{(slit Poiseuille).}
    \label{eq:stresspoise}
  \end{align}
\end{subequations}
where $\hat \sigma := \epsilon \hat{\dot{\gamma}} + \hat S$ is the
total stress.
 
The constitutive equations read
\begin{subequations}
  \label{eq:RDpoise}
  \begin{align}
    \partial_{\hat t}\hat S &=\hat D \partial^2_{\hat r}\hat S - \hat
    S + \hat{\dot{\gamma}} (1- \hat W)  \\
    \partial_{\hat t}\hat W &= \hat D \partial^2_{\hat r}\hat W -
    \hat W + \hat{\dot{\gamma}} \hat S  \\
    \partial_{\hat t}\hat Z &= \hat D \partial^2_{\hat r}\hat Z - \hat
    Z,
  \end{align}
\end{subequations}
and the shear rate condition is:
\begin{equation}
  \hat V = \int_0^1 \hat{\dot{\gamma}}(\hat r) d \hat r.
  \label{eq:sr1}
\end{equation}
\subsection{Cylindrical Couette flow}
In the Couette geometry with concentric cylinders at radii $R_1<R_2$
and assuming circular stream lines we have ${\bf v}=v(r) {\bf
  e}_{\theta}$.  The momentum balance in cylindrical coordinates reads
$\alpha \partial_{\hat t}\hat v= \frac{1}{\hat{r}^2} \partial_{\hat
  r}[{\hat r}^2(\hat S + \epsilon \hat{\dot{\gamma}} )]$, where
$\hat{\dot{\gamma}} = \hat{r} \frac{\partial}{\partial \hat{r}} \left(
  \frac{\hat v}{\hat r} \right)$ is the shear rate and $\alpha =
\frac{\rho R_1^2}{\mu \tau}$ has the same significance as in planar
shear.  Creeping flow corresponds to:
\begin{equation}
  \epsilon \hat{\dot{\gamma}} + \hat S = \hat \sigma (\hat r) =
  \frac{\hat \Gamma}{{\hat r}^2}. 
  \label{eq:stresscouette}
\end{equation}

The constitutive equations read
\begin{subequations}
  \label{eq:RDcouette}
  \begin{align}
    \partial_{\hat t}\hat S &=\hat D \Delta_{\hat r}\hat S - \hat S
    \left[1 + \frac{4 \hat D}{(1-a^2){\hat r}^2}\right]
    + \hat{\dot{\gamma}} (1- \hat W)  \\
    \partial_{\hat t}\hat W &= \hat D \Delta_{\hat r}\hat W - \hat W +
    \left[\hat{\dot{\gamma}} - \frac{4 \hat D}{(1-a^2){\hat
          r}^2}\right]\hat S  \\
    \partial_{\hat t}\hat Z &= \hat D \Delta_{\hat r}\hat Z - \hat
    Z\left[1-\frac{4 \hat D a^2}{(1-a^2){\hat r}^2}\right] + \frac{4
      \hat D a}{(1-a^2){\hat r}^2} \hat W
  \end{align}
\end{subequations}
where $\Delta_{\hat r} = \partial^2_{\hat r} + \frac{1}{\hat
  r}\partial_{\hat r}$ is the Laplacian.  The shear rate condition is:
\begin{equation}
  \hat V = \int_1^{1+p} \hat{\dot{\gamma}}(\hat r) \frac{d \hat r}{ \hat r} 
  \label{eq:sr2}
\end{equation}
where $p=\frac{R_2-R_1}{R_1}$.

All the above variables are dimensionless and have the same sign, with
the following definitions: $\hat V = - V
\frac{\tau}{R_1}\sqrt{1-a^2}$, $\hat W = (\frac{1-a}{2} \Sigma_{\theta
  \theta} - \frac{1+a}{2}\Sigma_{rr}) \frac{\tau}{\mu}$, $\hat Z =
(\frac{1-a}{2} \Sigma_{\theta \theta} + \frac{1+a}{2}\Sigma_{rr})
\frac{\tau}{\mu}$, $\hat S = \Sigma_{r \theta}
\frac{\tau}{\mu}\sqrt{1-a^2}$, ${\hat D\/} = {\cal D\/}
\frac{\tau}{R_1^2}$, $\hat \Gamma = - \Gamma \frac{\tau
  \sqrt{1-a^2}}{\mu R_1^2}$, $\hat \sigma = T_{r\theta}
\frac{\tau}{\mu}\sqrt{1-a^2}$, and $\hat r = r / R_1$.  $\Gamma$ is
the torque per unit length applied at the inner cylinder, and has the
same sign as $V$, the velocity of the inner cylinder, being opposite
to the shear rate and stress.  With these definitions $\hat \Gamma$
and $\hat V$ have the same sign as the shear rate and stress.
\subsection{Pipe flow}
Like for Couette flow we use cylindrical coordinates, with the z axis
parallel to the stream lines, ${\bf v}=v(r){\bf e}_z$.  The pipe is an
infinite length cylinder of radius $R$.  The momentum balance reads
$\alpha \partial_{\hat t}\hat v= \frac{1}{\hat{r}} \partial_{\hat
  r}[{\hat r}(\hat S + \epsilon \hat{\dot{\gamma}} )] - 2f$, where
$\hat{\dot{\gamma}} = \frac{\partial \hat v}{\partial \hat r}$ is the
shear rate.  Creeping flow corresponds to a total stress that is
linear in $\hat r$, like for the slit geometry:
\begin{equation}
\epsilon \hat{\dot{\gamma}} + \hat S = \hat \sigma (\hat r) = f \hat r.
\label{eq:creep_pipe}
\end{equation}
The shear rate condition is:
\begin{equation}
  \hat V = \int_0^1 \hat{\dot{\gamma}}(\hat r) d \hat r.
  \label{eq:sr1bis}
\end{equation}

The constitutive equations are:
\begin{subequations}
  \label{eq:RDpipe}
  \begin{align}
    \partial_{\hat t}\hat S &=\hat D \Delta_{\hat r}\hat S - \hat S
    \left[1 + \frac{ \hat D}{{\hat r}^2}\right]
    + \hat{\dot{\gamma}} (1- \hat W)  \\
    \partial_{\hat t}\hat W &= \hat D \Delta_{\hat r}\hat W - \hat W
    + \hat{\dot{\gamma}}\hat S  - \frac{4 \hat D \hat X}{{\hat r}^2 }  \\
    \partial_{\hat t}\hat Z &= \hat D \Delta_{\hat r}\hat Z -
    \hat Z + \frac{\hat D \hat X}{{\hat r}^2}  \\
    \partial_{\hat t}\hat X &= \hat D \Delta_{\hat r}\hat X + \hat X
    \left[ 1 + \frac{4 \hat D}{{\hat r}^2} \right] -
    \hat{\dot{\gamma}}\hat S
  \end{align}
\end{subequations}
We have used the following rescalings: $\hat V = - V_z
\frac{\tau}{R}\sqrt{1-a^2}$, $\hat W = (\frac{1-a}{2} \Sigma_{z z} -
\frac{1+a}{2}\Sigma_{rr}) \frac{\tau}{\mu}$, $\hat X = (\Sigma_{\theta
  \theta} - \Sigma_{z z}) \frac{\tau (1 + a)}{\mu}$, $\hat Z =
(\frac{1-a}{2} \Sigma_{z z} + \frac{1+a}{2}\Sigma_{rr})
\frac{\tau}{\mu}$, $\hat S = \Sigma_{r z}
\frac{\tau}{\mu}\sqrt{1-a^2}$, $f = \frac12\frac{\partial p}{\partial
  z}\frac{R\sqrt{1-a^2}}{\mu \tau}$, ${\hat D\/} = {\cal D\/}
\frac{\tau}{R^2}$, and $\hat r = r / R$.
\section{Matched asymptotic solution for the steady flow}\label{sect:matched}
The flow is described by the system of nonlinear, parabolic partial
differential equations of the reaction-diffusion type,
Eqs.~(\ref{eq:RDpoise}), (\ref{eq:RDcouette}), or (\ref{eq:RDpipe}).
The nonlinear reaction terms are due to the polymer stress relaxation.
The non-affine deformation (slip) is essential for the nonlinearity,
and all rescalings are possible for $|a| < 1$.

We consider small $\hat D$, for which the diffusion terms represent a
small perturbation to the steady flow equations.  Nevertheless, this
perturbation is singular and the solution can not be represented as a
uniformly convergent power series in $\hat D$.  The usual technique
applying to this situation is asymptotic matching \cite{lagerstrom88}.
The solution is divided into an inner layer solution around the
interface where the diffusion terms are important and outer layer
solutions a distance from the interface farther than its width (that
scales like $\hat D^{1/2}$), where diffusion terms are exponentially
small and can be neglected.
\subsection{Outer layer solution}
Neglecting terms of order $\hat D$ and imposing stationarity in any of
either Eqs.~(\ref{eq:stressplan}) and (\ref{eq:RDpoise}), or
Eqs.~(\ref{eq:stresscouette}) and (\ref{eq:RDcouette}), or
Eqs.~(\ref{eq:creep_pipe}) and (\ref{eq:RDpipe}), depending on
geometry, we obtain the same equations:
\begin{subequations}
  \label{eq:local}
  \begin{align}
    \hat{\dot{\gamma}}_{out} &=
    \frac{\hat \sigma(\hat r) - {\hat S}_{out}}{\epsilon}
    \label{eq:s1}\\
    {\hat S}_{out}&= \frac{\hat \sigma(\hat r) - {\hat
    S}_{out}}{\epsilon} (1 - {\hat W}_{out}) \label{eq:s2}\\ 
    {\hat W}_{out}&= {\hat S}_{out} \frac{\hat \sigma(\hat r) - {\hat
    S}_{out}}{\epsilon}\label{eq:s3}  \\ 
    {\hat Z}_{out}&= 0.\label{eq:s4}
  \end{align}
\end{subequations}
For pipe flow we may also use $\hat X_{out}=\hat{W}_{out}$ that
together with Eq.~(\ref{eq:s4}) implies $\left( \Sigma_{\theta,\theta}
\right)_{out}=0$.

\begin{figure}[tbh] \displaywidth\columnwidth \epsfxsize=5.5truein
  \centerline{{\epsfbox{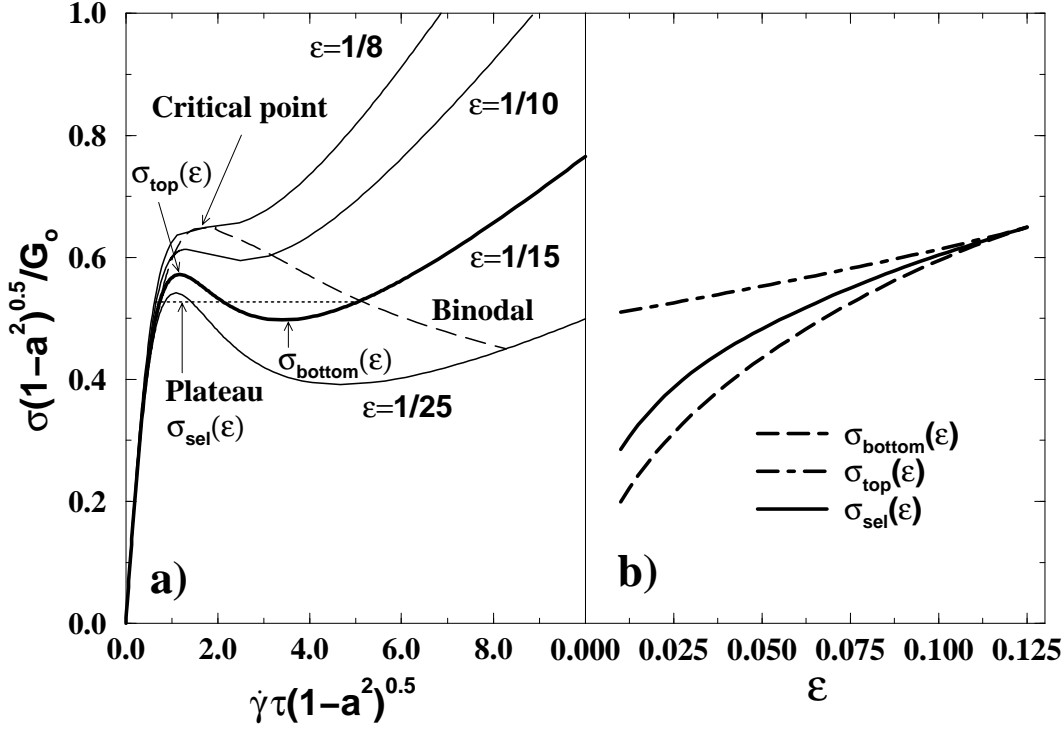}}}
  \caption{a) Local constitutive curves in reduced variables, 
    for different retardation parameters $\epsilon$, analogous to the
    van der Waals liquid-gas isotherms. The extremities of the stress
    plateau, for different $\epsilon$, form the binodal curve. Above
    $\epsilon=1/8$ (critical point) the constitutive curve is
    monotonic.  The homogeneous branches and the plateau constitute
    the flow curve in planar shear.  b) Limiting values of the stress
    allowing coexistence of bands, as functions of the retardation
    parameter $\epsilon$. $\sigma_{sel}$ is the selected stress value
    at the interface.  The values are consistent with the results in
    \protect\cite{jsplanar} obtained with a different method.  }
  \label{fig:localconst}
\end{figure}

The algebraic system of Eqs.~(\ref{eq:local}) is parametrised by $\hat
\sigma$, the value of the total stress at the position $\hat r$ in the
gap. The shear rate $\hat{\dot{\gamma}}$, as a function of the total
stress $\hat \sigma$ represents the local constitutive curve shown in
Fig.~\ref{fig:localconst}a. Banded flow is possible between the
minimum ($\hat \sigma = \sigma_{bottom}(\epsilon)$) and the maximum
($\hat \sigma = \sigma_{top}(\epsilon)$) of the flow curve, when for
one value of the stress there are three possible solutions of
Eqs.~(\ref{eq:local}).  Only two of those (low
$\{\hat{\dot{\gamma}}^{-}(\hat \sigma), {\hat S}^{-}(\hat \sigma),
{\hat Z}^{-}(\hat \sigma)\}$ and high shear rate
$\{\hat{\dot{\gamma}}^{+}(\hat \sigma), {\hat S}^{+}(\hat \sigma),
{\hat Z}^{+}(\hat \sigma)\}$) correspond to stable bands, the
intermediate shear rate being unstable \cite{malkus90,malkus91}. The
dependence of $\sigma_{bottom}$ and $\sigma_{top}$ on $\epsilon$ can
be obtained from Eqs.~(\ref{eq:local}), and is represented in
Fig.~\ref{fig:localconst}b.
\subsection{Inner layer solution}
In any of the Eqs.~(\ref{eq:stressplan}) and (\ref{eq:RDpoise}), or
Eqs.~(\ref{eq:stresscouette}) and (\ref{eq:RDcouette}), or
Eqs.~(\ref{eq:creep_pipe}) and (\ref{eq:RDpipe}), we impose
stationarity and change variables to:
\begin{equation}
  \label{eq:1}
  \tilde{r}=\frac{\hat r - {\hat r}^*}{\sqrt{\hat D}}
\end{equation}
where ${\hat r}^*$ is the position of the interface.

The stress is expanded about the interface position, $\hat \sigma =
\hat{\sigma}^* + {\hat D}^{1/2}\tilde{r} \partial_{\hat r}\hat \sigma
+ {\cal O\/}({\hat D})$, where $\hat{\sigma}^* =
\hat{\sigma}(\hat{r}^*)$ is the value of the stress at the interface.
By neglecting terms of order $\hat{D}^{1/2}$ and higher we obtain, for
all studied geometries:
\begin{subequations}
  \label{eq:inner}
  \begin{align}
    \partial^2_{\tilde{r}}{\hat S}_{in} &= {\hat S}_{in} -
    \frac{\hat{\sigma}^* -
      {\hat S}_{in}}{\epsilon} ( 1 - {\hat W}_{in} )  \\
    \partial^2_{\tilde{r}}{\hat W}_{in} &= {\hat W}_{in} - \frac{{\hat
        S}_{in}(\hat{\sigma}^* -
      {\hat S}_{in})}{\epsilon} \\
    {\hat Z}_{in} &= 0
    \label{eq:zinzero}
  \end{align}
\end{subequations} 
For pipe flow we may also use $\hat X_{in}= \hat{W}_{in}$, that
together with Eq.~(\ref{eq:zinzero}) implies $\left(
  \Sigma_{\theta,\theta} \right)_{in}=0$.

We refer here only to single interface solutions. More complex
solutions with an arbitrary number of interfaces can be treated in the
same way.  There are two types of single interface solutions,
differing by the sequence of bands. Let us refer to the solution with
the high shear rate band at the left of the interface (towards smaller
$\hat r$) as $(+-)$, and to the solution with the high shear rate band
at the right of the interface (towards bigger $\hat r$) as $(-+)$.

The inner and outer solutions should match at the interface (Prandtl
matching principle, Ref.~\cite{lagerstrom88}), leading for the
sequence $(+-)$ to:
\begin{subequations}
  \label{eq:prandtl1}
  \begin{align}
    {\hat S}_{in}^{+-}(-\infty) &= {\hat
      S}_{out}^{+}(\hat{\sigma}^{*}) & 
    {\hat W}_{in}^{+-}(-\infty) &= {\hat
      W}_{out}^{+}(\hat{\sigma}^{*}) \\
    {\hat S}_{in}^{+-}(\infty) &= {\hat
      S}_{out}^{-}(\hat{\sigma}^{*}) & 
    {\hat W}_{in}^{+-}(\infty) &= {\hat W}_{out}^{-}(\hat{\sigma}^{*}) 
    \intertext{or for the sequence $(-+)$ to:}
    {\hat S}_{in}^{-+}(-\infty) &= {\hat S}_{out}^{-}(\hat{\sigma}^{*}) & 
    {\hat W}_{in}^{-+}(-\infty) &= {\hat W}_{out}^{-}(\hat{\sigma}^{*})  \\
    {\hat S}_{in}^{-+}(\infty) &= {\hat S}_{out}^{+}(\hat{\sigma}^{*}). &
    {\hat W}_{in}^{-+}(\infty) &= {\hat W}_{out}^{+}(\hat{\sigma}^{*})
  \end{align}
\end{subequations}
\subsection{Equations of flow curves}
The flow curve expresses the relation between measurable quantities:
velocity of the inner cylinder $\hat V$ and torque $\hat \Gamma$ in
the Couette geometry, maximum velocity $\hat V$ and pressure gradient
$f$ in Poiseuille flow, gap velocity $\hat V$ and shear stress $\hat
\sigma$ in planar shear.
 
In order to obtain the flow curves we use the shear rate conditions,
Eqs.~(\ref{eq:sr1}), (\ref{eq:sr2}), (\ref{eq:sr1bis}), and the
relations (\ref{eq:stresspoise}), (\ref{eq:stresscouette}),
(\ref{eq:creep_pipe}) between the total stress $\hat \sigma$ and the
position $\hat r$.  In the banded regime, the shape of the flow curve
depends on the number of interfaces and on the sequence of bands.  We
shall show in the next section that in the presence of stress
diffusion steady flow always corresponds or can be reduced by symmetry
to a single interface.

In Poiseuille flow, using Eqs.~(\ref{eq:stresspoise}), (\ref{eq:sr1})
or (\ref{eq:creep_pipe}), (\ref{eq:sr1bis}), a single interface flow
curve with the high shear rate band near the wall (right of the
interface) is described by:
\begin{equation}
  {\hat V}(f,\hat{\sigma}^*)=
  \frac{1}{f}\left[ \int_{0}^{\hat{\sigma}^*}
    \hat{\dot{\gamma}}^-(\hat \sigma)d\hat \sigma +  
  \int_{\hat{\sigma}^*}^{f} \hat{\dot{\gamma}}^+(\hat \sigma)d\hat
  \sigma \right] 
  \label{eq:v-+poise}
\end{equation}
where $\hat{\sigma}^* = \hat{\sigma} (\hat{r}^*)$ is the value of the
total stress at the position of the interface.  The inverse sequence
of bands is prohibited in Poiseuille flow, because the high shear rate
band can not exist at $\hat r =0$, where $\hat \sigma =0$.

In Couette flow, using Eqs.~(\ref{eq:stresscouette}), (\ref{eq:sr2}) a
single interface flow curve with the high shear rate band near the
inner cylinder (left of the interface) is given by:
\begin{equation}
  {\hat V}^{+-}(\hat \Gamma,\hat{\sigma}^*)=
  \frac{1}{2}\left[ \int_{\hat{\Gamma}/(1+p)^2}^{\hat{\sigma}^*}
    \hat{\dot{\gamma}}^-(\hat \sigma)     \frac{d\hat \sigma}{\hat \sigma} + 
    \int_{\hat{\sigma}^*}^{\hat \Gamma} \hat{\dot{\gamma}}^+(\hat \sigma)
    \frac{d\hat \sigma}{\hat \sigma} \right]
  \label{eq:v+-couette}
\end{equation}

The flow curve for the inverse sequence of bands (high shear rate band
near the outer cylinder) is given by:
\begin{equation}
  {\hat V}^{-+}(\hat \Gamma,\hat{\sigma}^*)=  \frac{1}{2}\left[
    \int_{\hat{\Gamma}/(1+p)^2}^{\hat{\sigma}^*} 
    \hat{\dot{\gamma}}^+(\hat \sigma) 
    \frac{d\hat \sigma}{\hat \sigma} + 
    \int_{\hat{\sigma}^*}^{\hat \Gamma} \hat{\dot{\gamma}}^-(\hat \sigma)
    \frac{d\hat \sigma}{\hat \sigma} \right]
  \label{eq:v-+couette}
\end{equation} 

Planar flow represents a special case because the total stress is
constant throughout the gap (Eq.~(\ref{eq:stressplanar})) and thus:
\begin{equation}
  \hat V (\nu ,\hat{\sigma}) = \nu \hat{\dot{\gamma}}^+(\hat{\sigma}) + 
  (1 - \nu )\hat{\dot{\gamma}}^-(\hat{\sigma})
  \label{eq:vplanar}
\end{equation}
where $\nu$ is the proportion of high shear rate band in the flow.  As
we shall argue below, the flow curves above are well defined because
the value $\hat{\sigma}^*$ of the shear stress at the interface is a
geometry-independent constant.
\section{Uniqueness and stability of the steady flow}
\label{sect:uniqueness}
\subsection{An existence conjecture for the inner layer solution}

The inner layer solution controls the existence and stability of
steady banded flows.  Because we are dealing with singular
perturbations, this control can be performed even by a very thin
interface (very small $\hat D$).  Before stating an important property
of the inner layer solution, we note that steady banded flow is a
particular case of a moving interface solution $\hat{r}^*(t)$. We may
look for matched asymptotic solutions in this case as well and one may
show by using the change of variable
\begin{equation}
  \tilde{r}=\frac{\hat r - {\hat r}^*(t)}{\sqrt{\hat D}}
\end{equation}
that the moving interface inner layer solution should obey:
\begin{subequations}
  \label{eq:movinginterface}
  \begin{align}
    \partial^2_{\tilde{r}}{\hat S}_{in} +
    c \partial_{\tilde{r}}{\hat S}_{in} &= {\hat S}_{in} - \frac{\hat
      \sigma^* -  {\hat S}_{in}}{\epsilon} ( 1 - {\hat W}_{in} )  \\
    \partial^2_{\tilde{r}}{\hat W}_{in} + c \partial_{\tilde{r}}{\hat
      W}_{in} &= {\hat W}_{in} - \frac{{\hat S}_{in}(\hat \sigma^* - 
      {\hat S}_{in})}{\epsilon}
  \end{align}
\end{subequations}
where
\begin{equation}
  c= \frac{1}{\sqrt{\hat D}}\frac{d \hat{r}^*(t)}{d t}
  \label{eq:rescaled_c}
\end{equation}
is the rescaled velocity of the interface.\\

\begin{cnj}
  There is a unique value $\hat{\sigma}^*=\hat{\sigma}_{sel} \in
  (\sigma_{bottom}(\epsilon),\sigma_{top}(\epsilon))$ of the total
  shear stress at the position of the interface, such that a steady
  inner layer solution for the banded flow of the JS-d model
  (Eqs.~(\ref{eq:inner}) exists obeying the Prandtl matching
  principle, Eqs.~(\ref{eq:prandtl1}). Moreover, moving interface
  inner layer solutions obeying the Prandtl matching principle for
  Eqs.~(\ref{eq:movinginterface}) exist for values of $\hat{\sigma}^*$
  in a neighbourhood of $\hat{\sigma}_{sel}$, with velocities
  $c(\hat{\sigma}^*)$ that are well defined functions of
  $\hat{\sigma}^*$.  The stationary solution satisfies
  $c(\hat{\sigma}_{sel}) = 0$ and, furthermore, $\frac{d c}{d
    \hat{\sigma}^*}|_{\hat{\sigma}_{sel}} > 0 $ when the high shear
  rate band is at the left (towards smaller values of $\hat r$) of the
  interface, and $\frac{d c}{d \hat\sigma^*}|_{\hat{\sigma}_{sel}} < 0
  $ for the inverse sequence of bands.\\
\end{cnj}

\begin{itemize}
\item []
\begin{rmk}
  The conjecture considers the inner layer solution, which obeys the
  same equation for all flow geometries. This implies that the
  selected value of the stress is geometry-independent.  After the
  rescalings, the diffusion coefficient does not enter
  Eqs.~(\ref{eq:movinginterface}). Thus, the selected stress is
  independent of $\cal D$. As shown by Eq.~(\ref{eq:rescaled_c}) the
  speed of the interface $d\hat{r}^{\ast}/dt$ scales with ${\hat
    D}^{1/2}$, thus the
  r\^{o}le of $\hat D$ is to speed up kinetics of non-steady flow.\\
\end{rmk}

\begin{rmk}
  A stationary inner layer solution which obeys the Prandtl matching
  principle represents a heteroclinic orbit of the 4D dynamical system
  $\left\{\hat S, \partial_{\tilde r}\hat S\right.$, $\left.  \hat
    W,\partial_{\tilde r}\hat W \right\}$, connecting the hyperbolic
  fixed points $\left\{\hat{S}^+(\hat{\sigma}_{sel})\right.$, $0$,
  $\hat{W}^+(\hat{\sigma}_{sel})$, $0 \bigr\}$ and
  $\left\{\hat{S}^-(\hat{\sigma}_{sel}), 0 ,
    \hat{W}^-(\hat{\sigma}_{sel}), 0 \right\}$.  As discussed in
  \cite{olmradlu99,jsplanar}, and as is valid rather generally in
  these cases, the value of the parameter
  $\hat{\sigma}^*=\hat{\sigma}_{sel}$ allowing this solution is
  isolated (actually unique in the interval
  $(\sigma_{bottom}(\epsilon)$, $\sigma_{top}(\epsilon))$).  The
  condition on the sign of the derivative $\frac{d c(\hat \sigma)}{d
    \hat{\sigma}^*}$ means that an increase of the stress at the
  interface above $\hat{\sigma}_{sel}$ produces a movement of the
  interface that decreases the size of the low shear rate band.
  Conversely, a decrease of the stress at the interface below
  $\hat{\sigma}_{sel}$ produces a displacement of the interface that
  increases the size of the low shear rate band. The same property can
  be found for other reaction-diffusion systems ({\sl e.g.\/} the
  FitzHugh-Nagumo model of nerve conduction in biophysics) and has
  been occasionally referred to as ``dominance principle''
  \cite{bode97,comment}.\\
\end{rmk}

\begin{rmk}
  We have numerically checked this conjecture for various values of
  the parameter $\epsilon$ of the JS-d model (see next Section).  All
  of these features are easy to prove for toy models that lead to
  integrable dynamical systems \cite{eurorheo}.\\
\end{rmk}
\end{itemize}

\subsection{Numerical test of the conjecture}
In order to test the conjecture for the JS-d model, and determine the
relation between $c$ and $\hat{\sigma}^*$ for different values of the
unique parameter $\epsilon$ of the model in reduced variables, we have
numerically integrated the following system of partial differential
equations:
\begin{subequations}
  \label{eq:numsystem} 
  \begin{align}
    \partial_{t}S&=\beta \partial^2_{\hat{r}}S + c \sqrt{\beta}
    \partial_{\hat{r}}S - S + \frac{\sigma(\hat{r}) - S}{\epsilon} ( 1 - W ) \\
    \partial_{t}W&=\beta \partial^2_{\hat{r}}W + c \sqrt{\beta}
    \partial_{\hat{r}}W - W + \frac{S(\sigma(\hat{r}) - S)}{\epsilon}
    \intertext{with} 
    S(L)&=S^-(\sigma_{bottom}(\epsilon)),\quad
    S(-L)=S^+(\sigma_{top}(\epsilon)) \\
    W(L)&=W^-(\sigma_{bottom}(\epsilon)),\quad
    W(-L)=W^+(\sigma_{top}(\epsilon)),
  \end{align}
\end{subequations} 
where
\begin{equation}
  \sigma (\hat{r}) = \sigma_{top} (\epsilon)
  + (\sigma_{bottom}(\epsilon) - \sigma_{top}(\epsilon))\frac{r+L}{2L}
\end{equation}
is conveniently chosen to scan the interval
$[\sigma_{bottom}(\epsilon),\sigma_{top}(\epsilon)]$. Provided that
the length $L$ is taken much larger than the interface width, {\em
  i.e.} $\sqrt{\beta} \ll L$, the stationary solution is an interface
in a position corresponding to a value of the stress equal to
$\sigma^*=\hat{\sigma}^*(c)$.  $\beta$ can be interpreted as the diffusion
coefficient and $\sigma(\hat{r})$ as the total stress inside the gap
of width $2L$ although the applicability of the method does not rely
on its connection to a concrete physical problem, but on its formal
analogy to Eqs.~(\ref{eq:creep_plan}) and (\ref{eq:RDpoise}), in the
asymptotic $(L/\sqrt{\beta}\rightarrow\infty)$ limit.

In this way we determine
$\sigma_{sel}(\epsilon)=\sigma^*(c=0;\epsilon)$ and $\frac{d c}{d
  \sigma^*}(\sigma_{sel};\epsilon)= \left( \frac{d \sigma}{d c}
\right)^{-1}(c=0;\epsilon)$.  The first dependence will be used in
calculating the flow curves in Section \ref{sect:applications}, while
the sign of the latter constitutes a proof of the dominance principle.

\begin{figure}[tbh] \displaywidth\columnwidth \epsfxsize=5.0truein
  \centerline{{\epsfbox[40 25 733 410]{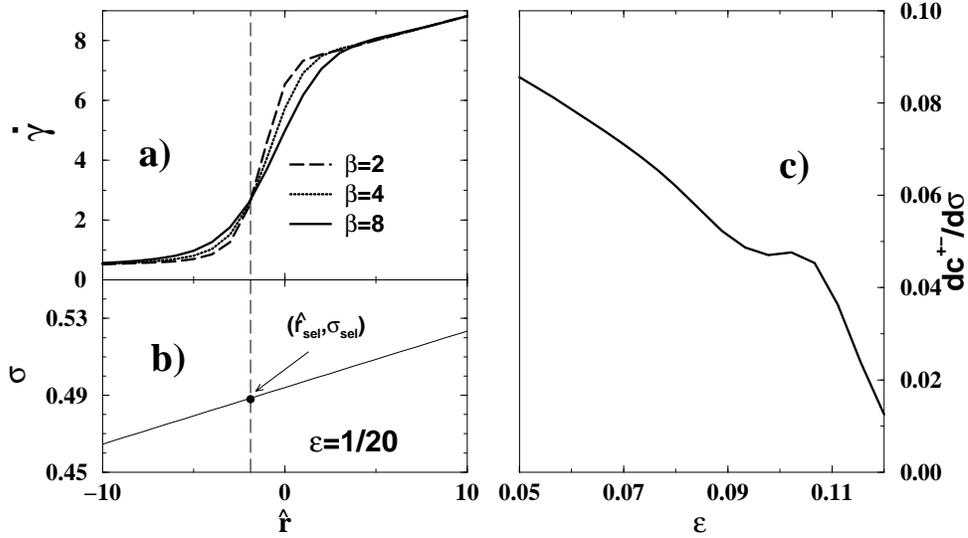}}}
\caption{a) Strain rate $\dot{\gamma}$ as a function of position for
  three different stationary solutions $(c=0)$ to
  Eq.~(\ref{eq:numsystem}) for $\beta=2,4,8$. (b) Stress as a function
  of position.  The cross-over of several stationary profiles gives
  the selected stress at the position noted by the vertical line. c)
  Derivative at $c^{+-}=0$ of $c^{+-}(\hat{\sigma}^*)$ as a function
  of the retardation parameter $\epsilon$, for an interface with the
  high shear rate band at its left. Here, $\sigma_{sel}=0.4799$,
  compared to $\sigma_{sel}=0.4827$ found in \cite{jsplanar} for the
  JS model in planar Couette flow by a different method.}
\label{fig:cross}
\end{figure}

In order to determine precisely the position of the stationary
interface and hence the selected stress, we have solved
Eqs.~(\ref{eq:numsystem}) for several values of $\beta$.  Changing
$\beta$ modifies the interface width, analogous to a change of the
diffusion coefficient, but does not alter the selected position of the
interface for small $\beta$.  The selected position of the interface
and the corresponding value of the stress are given by the cross-over
of several profiles obtained for different values of $\beta$ as in
Figs.~\ref{fig:cross}a,b.

This algorithm is more convenient than the one used in \cite{jsplanar}
that considers a constant stress $\sigma$ throughout the gap (planar
shear) and tunes ``by hand'' the value of $\sigma$ to obtain a
stationary interface. Here the tuning of the stress is automatically
obtained while integrating Eqs.~(\ref{eq:numsystem}).

The selected value of the stress is compared with $\sigma_{bottom}$
and $\sigma_{top}$ in Fig.~\ref{fig:localconst}b and agrees with the
result of the method in \cite{jsplanar}.

In order to test the dominance principle we have calculated
$\frac{dc}{d\hat{\sigma}^*}$.  Changing $\hat{r}\rightarrow-\hat{r}$ and
$c\rightarrow-c$ leaves Eqs.~(\ref{eq:numsystem}) invariant.  Thus if
$c^{+-}(\hat{\sigma}^*)$ and $c^{-+}(\hat{\sigma}^*)$ are the
velocities of the interface with the high shear rate at its left and
right, respectively, then
$c^{-+}(\hat{\sigma}^*)\,=\,-\,c^{+-}(\hat{\sigma}^*)$.  Hence, our
check that $\frac{dc^{+-}}{d\hat{\sigma}^*}>0$, as shown in
Fig.~\ref{fig:cross}c, represents the sufficient test of the
dominance principle.
\subsection{Number of bands and stability of banded flow}
The numerically tested conjecture implies that the interface is
stationary in a position inside the gap where the value of the stress
is $\hat{\sigma}_{sel}$.  This position is unique ($\hat{r}_{sel} =
(\hat{\Gamma}/\hat{\sigma}_{sel})^{1/2} $ for Couette flow,
$\hat{r}_{sel} = \hat{\sigma}_{sel}/f$ for pipe flow) because in
steady flow the total stress is monotonically dependent on the radius.
Therefore, in Couette and pipe geometries banded flow has only one
interface separating two bands.  For Poiseuille slit flow, the
dependence of the total stress on position is symmetric with respect
to the middle of the slit, so there are two interfaces at
$\hat{r}_{sel} = \pm \hat{\sigma}_{sel}/f $.

For both Poiseuille slit and pipe flow the sequence of bands is
unique, with a centre low shear rate band and outer high shear rate
bands (or an annular band in pipe flow).  In Couette flow, two
sequences of bands are possible under certain conditions that we shall
discuss in the next section.  There are no restrictions on the
sequence and order of the bands in planar flow because the total
stress is constant throughout the gap; interfaces can exist anywhere,
and any number and sequence of bands is possible, provided that the
bands are much thicker than the interface width.

The stability of banded steady flow in presence of diffusion terms was
tested in the Couette geometry by numerically evolving the dynamical
equations \cite{olmradlu99} for both normal (high shear rate at the
inner cylinder) and inverted sequences of bands.  We have also proved
(Appendix \ref{appendix1}) the stability of different types of banded
solutions in Poiseuille and Couette flows with respect to
perturbations of the position of the interface.  A linear stability
analysis for 1D perturbations around a banded profile, as well as the
stability with respect to surface waves in 2D, involves lengthy
functional analysis and will not be discussed here. Linear stability
of banded flow for 1D perturbations in slit Poiseuille geometry was
shown in Ref.~\cite{brunovsky94} for a simplified model with only one
order parameter.  For the JS model (without diffusion) linear
stability of the banded flow with any finite number of bands was
proven in \cite{nohel90} for 1D perturbations in Poiseuille geometry,
and possible instabilities induced by surface waves were discussed in
\cite{renardy}.

Although linearly stable, the inverted sequence of bands is metastable
because the interfaces separating a nucleus of low shear rate band
within the high shear rate band will find themselves in a region of
stress lower than $\hat{\sigma}_{sel}$ and according to the dominance
principle will have non-zero velocities oriented such that the nucleus
will grow \cite{eurorheo}.  The same is true for a nucleus of high
shear rate band in the region where the total stress is higher than
$\hat{\sigma}_{sel}$.
\section{Applications of the results}
\label{sect:applications}
\subsection{Flow curves and extremities of branches}
Flow curves can be obtained from Eqs.~(\ref{eq:v-+poise}-\ref{eq:vplanar}).
These curves have several branches corresponding to homogeneous and
banded flow.  The extremities of different branches are fixed by the
following conditions:

\begin{enumerate}
\item  Homogeneous low shear rate flow is possible if 
  \begin{equation}
    0 < \hat{\sigma}(\hat{r}) < \hat{\sigma}_{top}(\epsilon), \forall \hat{r}
    \label{eq:homlow}
  \end{equation}
\item
  Homogeneous high shear rate flow is possible if
  \begin{equation}
    \hat{\sigma}(\hat{r}) > \hat{\sigma}_{bottom}(\epsilon), \forall \hat{r}
    \label{eq:homhigh}
  \end{equation}
\item Banded flow is possible if
  \begin{equation}
    \hat{\sigma}(\hat{r}^*_{sel}) = \hat{\sigma}_{sel}
    \label{eq:banded}
  \end{equation}
  for some position
  $\hat{r}^*_{sel}$ inside the flow.
\end{enumerate}
The sequence $(+-)$ is allowed if, in addition:
\begin{align}
0 < \hat{\sigma}(\hat{r}) < \hat{\sigma}_{top}(\epsilon), 
&\forall \hat{r},\quad  \hat{r} > \hat{r}^*_{sel}  \\
\hat{\sigma}(\hat{r}) > \hat{\sigma}_{bottom}(\epsilon), 
&\forall \hat{r}, \quad \hat{r} < \hat{r}^*_{sel}
\label{eq:banded+-}
\end{align}
while the sequence $(-+)$ is allowed if, in addition:
\begin{align}
0 < \hat{\sigma}(\hat{r}) < \hat{\sigma}_{top}(\epsilon), 
&\forall \hat{r}, \quad \hat{r} < \hat{r}^*_{sel}  \\
\hat{\sigma}(\hat{r}) > \hat{\sigma}_{bottom}(\epsilon), 
&\forall \hat{r}, \quad \hat{r} > \hat{r}^*_{sel}
\label{eq:banded-+}
\end{align}

\begin{figure}[tbh] \displaywidth\columnwidth \epsfxsize=5truein
    \centerline{\epsfbox{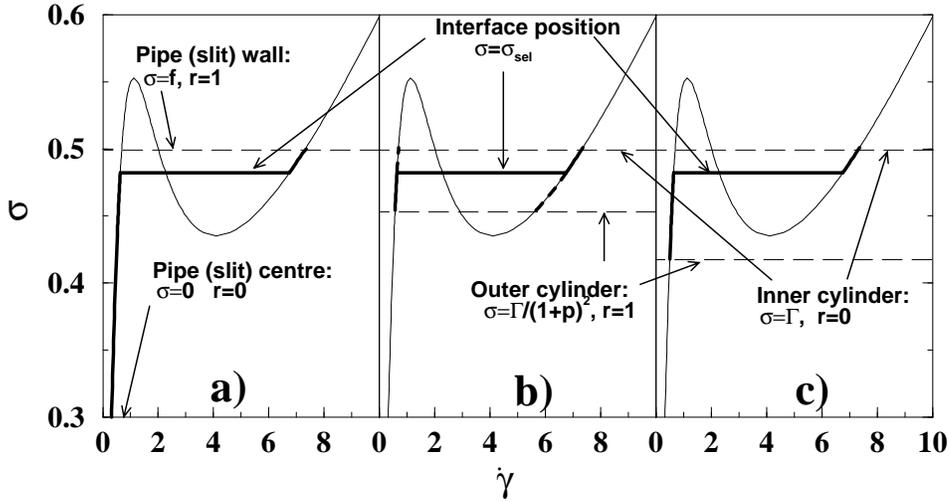}} 
\caption{Shear rate profiles (thick line) 
  for banded flow (to each value of the total stress $\sigma$
  corresponds a position $r$ inside the gap) in a) Poiseuille
  geometry, b) Couette geometry $p=0.05,\epsilon=1/20$ (region I of
  Fig.~\ref{fig:domains}), with two possible sequences of bands
  ($(+-)$ solid, high shear rate band at the inner cylinder), and
  $(-+)$ dashed, high shear rate band at the outer cylinder), c)
  Couette geometry $p=0.1,\epsilon=1/20$ (region IV of
  Fig.~\ref{fig:domains} inverted sequence is not possible).  The
  interface was considered of negligible width, hence it is
  represented by the segment $\sigma=\sigma_{sel}$.}
\label{fig:sigma_profiles}
\end{figure}
In Poiseuille and Couette geometries, any banded steady flow profile
can be conveniently represented as a segment on the local constitutive
curve $\hat{\dot{\gamma}}(\hat \sigma)$, a solution of
Eqs.~(\ref{eq:local}) for variable stress $\hat \sigma$
(Fig.~\ref{fig:sigma_profiles}).  Using the relation between $\hat r$
and $\hat \sigma$ (Eqs.~\ref{eq:stresspoise},\ref{eq:stresscouette},
\ref{eq:creep_pipe}) one can map a steady state
$\hat{\dot{\gamma}}(\hat r)$ onto a segment of the local constitutive
curve $\hat{\dot{\gamma}}(\hat \sigma)$ for stresses $0 < \hat{\sigma}
< f$ (Poiseuille), or $\hat{\Gamma}/(1+p)^2 < \hat{\sigma} <
\hat{\Gamma}$ (Couette). The outer layer solutions lies on the local
constitutive curve, while the inner layer solution is a horizontal
segment at $\hat{\sigma} = \hat{\sigma}_{sel}$, corresponding to a
small width interface.
\subsubsection{Planar shear}
The equation of the low shear rate branch is $\hat{V} =
\hat{\dot{\gamma}}^-(\hat{\sigma})$, with $0 < \hat{\sigma} <
\hat{\sigma}_{top}(\epsilon)$, and the high shear branch can be
described as $\hat{V} = \hat{\dot{\gamma}}^+(\hat{\sigma})$, with
$\hat{\sigma}_{bottom}(\epsilon) < \hat{\sigma}$. The banded flow
corresponds to the plateau $\hat{\sigma} = \hat{\sigma}_{sel}$. The
homogeneous branches of the 
flow curve are on 
the local constitutive curve represented in Fig.~\ref{fig:localconst}.
\subsubsection{Poiseuille  flow}
The low shear rate branch can be obtained from Eq.~(\ref{eq:v-+poise})
by choosing $\hat{\sigma}^* = f$ (interface near the wall at
$\hat{r}^* =1$), and according to Eq.~(\ref{eq:homlow}) with the
restriction $0 < f < \hat{\sigma}_{top}(\epsilon)$.  The unique banded
profile $(-+)$ (low shear rate at $\hat r = 0$) is described by
Eq.~(\ref{eq:v-+poise}), choosing $\hat{\sigma}^* =
\hat{\sigma}_{sel}$, according to Eq.~(\ref{eq:banded}) with the
restriction $ f= \hat{\sigma}_{sel}(\epsilon)/\hat{r}^*_{sel} >
\hat{\sigma}_{sel}(\epsilon)$. Then Eq.~(\ref{eq:banded-+}) is
automatically fulfilled because it becomes equivalent to
$\sigma_{bottom}(\epsilon) < \hat{\sigma}_{sel}(\epsilon) <
\sigma_{top}(\epsilon)$, that is true for $\epsilon < 1/8$ (actually
for all values of $\epsilon$ for which the local constitutive curve is
nomonotonic, Fig.~\ref{fig:localconst}). Eq.~(\ref{eq:banded+-}) is
not fulfilled (thus the inverted sequence of bands is forbidden)
because it becomes equivalent to $f\hat{r} > \sigma_{bottom}, \forall
\hat{r}<\hat{r}^*_{sel}$, obviously not true for $\hat r = 0$.  The
banded branch continues until $f = \infty$ and $\hat{r}^* = 0$. The
low shear rate band is continuously squeezed until it disappears in
the limit $f = \infty$. Thus, the flow curve has no high shear rate
branch, because the high shear rate band occupies the entire volume
only at $f = \infty$.  The flow curve in this geometry is represented
in Fig.~\ref{fig:flow_curves}a for different retardation parameters
$\epsilon$.
\subsubsection{Cylindrical Couette flow}
In Couette flow Eq.~(\ref{eq:banded+-}) is always fulfilled and the
normal sequence $(-+)$ of bands is allowed for all shear rates along
the plateau (see Fig.~\ref{fig:sigma_profiles}b-d). Nevertheless, the
inverted sequence $(-+)$ is not allowed for the values of $\Gamma, p,
\epsilon$ when the total stress at the inner cylinder is higher than
$\sigma_{top}$, or when the total stress at the outer cylinder is
lower than $\sigma_{bottom}$, see Fig.~\ref{fig:sigma_profiles}c. A
detailed analysis of this condition given in Appendix~\ref{appendix2}
shows the existence of five possibilities and regions in the plane of
parameters $(p,\epsilon)$, explained in Fig.~\ref{fig:domains}.
\begin{figure}[tbh]
  \displaywidth\columnwidth \epsfxsize=5truein
  \centerline{{\epsfbox{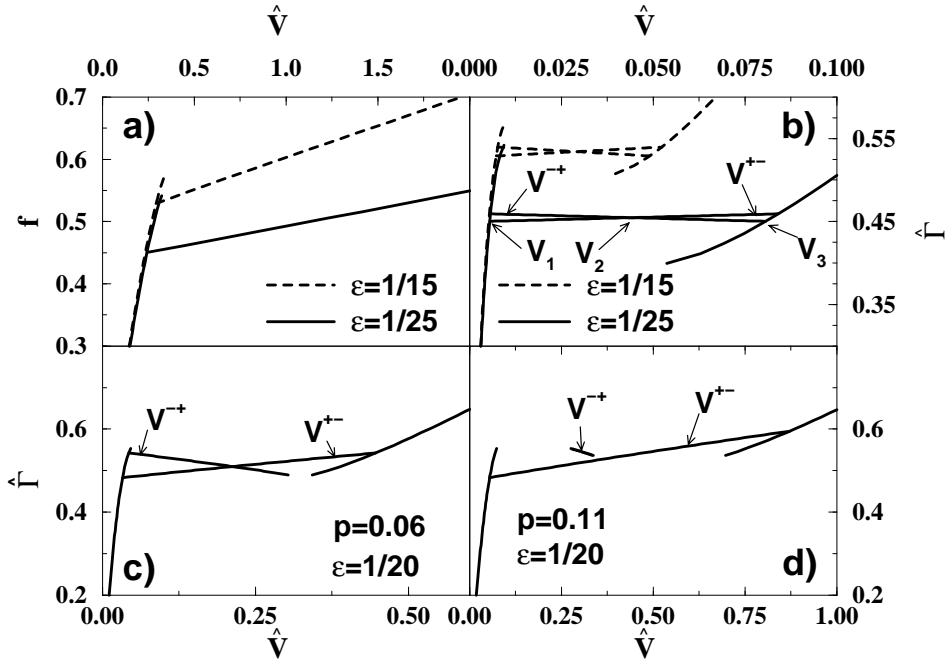}}}
\caption{Flow curves in reduced variables for a) Poiseuille flow, b-d)
  Couette geometry with $p=0.01$ (b), $p=0.06$ (c), and $p=0.11$ (d).
  For Couette flow, $V^{+-}$ and $V^{-+}$ correspond to flow curves
  with the high strain rate band near the inner and outer cylinders,
  respectively. (b) and (c) correspond to Regions I and II, while (d)
  corresponds to Region IV in Fig.~\protect{\ref{fig:domains}}.}
\label{fig:flow_curves}
\end{figure}
\begin{figure}[tbhe] \displaywidth\columnwidth \epsfxsize=5.0truein
  \centerline{\epsfbox{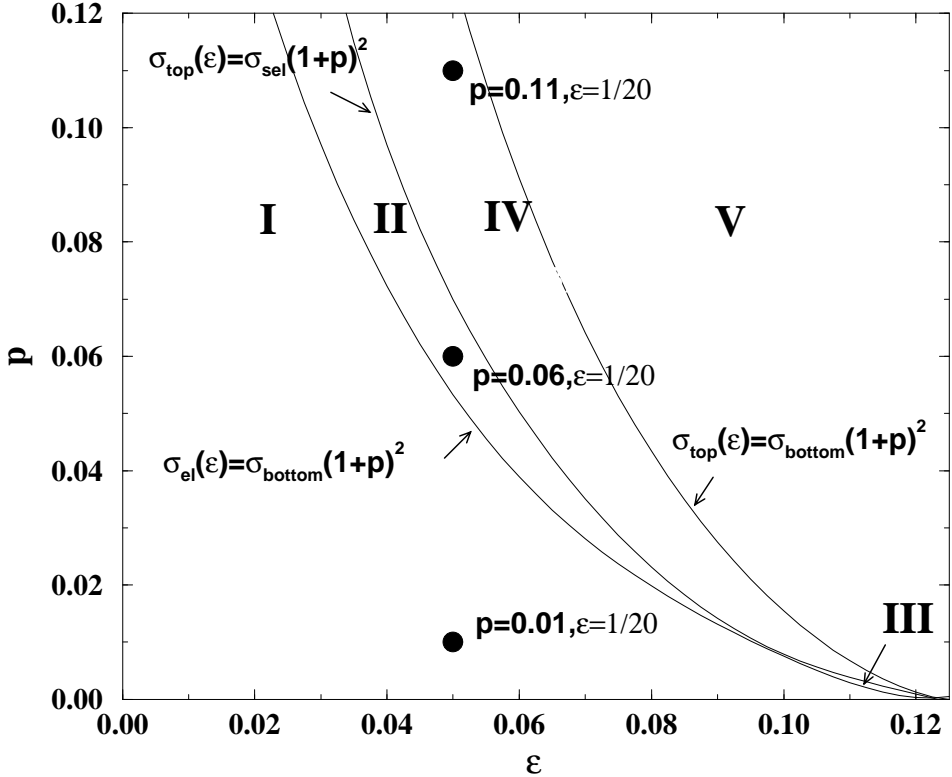}}
\caption{Existence conditions for the inverted sequence of bands
  in Couette flow, separating 5 regions I-V in the plane
  $(\epsilon,p)$. The dots $\bullet$ correspond to the flow curves in
  Fig.~\protect{\ref{fig:flow_curves}}. Inverted bands are only
  possible in Regions I (for all shear rates along the plateau), II
  (for a domain of shear rates truncated at the right end of the
  plateau), III (for a domain of shear rates truncated at the left end
  of the plateau), and IV (for a domain of shear rates truncated at
  both ends of the plateau). }
\label{fig:domains}
\end{figure}

Some examples of flow curves in the Couette geometry are presented in
Fig.~\ref{fig:flow_curves}.  Eq.~(\ref{eq:dv+-gamma}) implies that the
branch $(+-)$ has a positive slope $\frac{\partial
  \hat{\Gamma}}{\partial \hat{V}} > 0$, while Eq.~(\ref{eq:dv-+gamma})
implies that the branch $(-+)$ has a negative slope $\frac{\partial
  \hat{\Gamma}}{\partial \hat{V}} < 0$.  The latter does not imply a
mechanical instability, because any mechanical drift is constrained by
the imposed mean shear rate. Furthermore, the stability of this
sequence of bands is shown by our numerical simulations
\cite{olmradlu99} and by the stability analysis with respect to the
position of the interface (Appendix~\ref{appendix1}).  As suggested in
\cite{geovla98,fyrilas99}, stability of banded flow is not directly
connected to the slope of the flow curve, but rather to the slope of
the local constitutive curve of the homogeneous bands; if the latter
is negative for at least one band (intermediate branch in
Fig.~\ref{fig:localconst}a), then the flow is unstable.

Fig.~\ref{fig:flow_curves}b corresponds to values of the parameters
$(p,\epsilon)$ chosen in Region I of Fig.~\ref{fig:domains}, with both
sequences of bands allowed.  Fig.~\ref{fig:flow_curves}c corresponds
to parameters chosen in Region II (the inverted band sequence is
forbidden at higher strain rates), while Fig. \ref{fig:flow_curves}d
corresponds to parameters chosen in Region IV (shear rates allowing
the existence of the inverted sequence are limited both at upper and
lower values).  Interestingly, the branch corresponding to the
inverted band sequence no longer crosses the normal band sequence
branch for $p=0.11, \epsilon = 0.05$ in Fig.~\ref{fig:flow_curves}d.
Note that the ``plateau'' in the banded region is steeper (larger
$d\hat{\Gamma}/d\hat{V}$) for a more highly curved geometry (larger
$p$).

The various dangling segments corresponding to different branches of
the flow curves may be the origin of hysteresis phenomena.  Part of
these phenomena were described in a previous paper \cite{olmradlu99}
using numerical integration of the dynamical equations.  We have shown
there that some segments of the flow curve can be reached only by
special experimental scenarios. This is in agreement with the results
of this work.  For instance, starting from rest in
Fig.~\ref{fig:flow_curves}b, when $\epsilon = 1/25$, the system can
first reach the branch $(+-)$ within the interval of gap velocities $
\hat{V}_1 < \hat{V} < \hat{V}_2$ and the branch $(-+)$ within the
interval $\hat{V}_2 < \hat{V} < \hat{V}_3$
(Fig.~\ref{fig:flow_curves}b).  In order to reach other parts of the
banded branches one must, for example, prepare the system with a given
band sequence and then adiabatically change the value of the gap
velocity in order to scan the entire length of the branch.  This is
possible for $p < 0.1$ for both bands $(+-)$, but for $p=0.11$ in
Fig.~\ref{fig:flow_curves}d all start-up preparations of the flow
end with the sequence $(+-)$ and it is impossible to reach the
isolated branch $(-+)$.  Numerically one could start from rest with a
smaller value of $p$, reaching thus the $(-+)$ branch, and then change
$p$ adiabatically, but this is not a conceivable experimental
procedure.
\subsection{Interface Width}
Linearising Eqs.~(\ref{eq:inner}) at $\tilde{r}=\pm \infty$ we obtain
\begin{equation}
\partial^2_{\tilde{r}}\left(
        \begin{matrix}
                \delta \hat{S}_{in} \\
                \delta \hat{W}_{in}
        \end{matrix}
        \right)
        =  {\bf M}^{\pm} \left(
        \begin{matrix}
                \delta \hat{S}_{in} \\
                \delta \hat{W}_{in}
        \end{matrix}
        \right) 
\end{equation}
where $\delta \hat{S}_{in},\delta \hat{W}_{in}$ are small deviations
of stresses with respect to asymptotic steady values and
\begin{equation}
  {\bf M}^{\pm}(\hat{\sigma}_{sel}) = 
  \left( 
    \begin{matrix}
      1 - [ 1-\hat{W}^{\pm} (\hat{\sigma}_{sel}) ]/\epsilon & 
      [ \hat{\sigma}_{sel} - \hat{S}^{\pm} (\hat{\sigma}_{sel}) ] / \epsilon \\
      \protect [ \hat{\sigma}_{sel} - 2 \hat{S}^{\pm} (\hat{\sigma}_{sel}) ] / \epsilon & 1
    \end{matrix}
  \right)
  \label{eq:linear}
\end{equation}

\begin{figure}[tbh] \displaywidth\columnwidth \epsfxsize=5.0truein
    \centerline{{\epsfbox{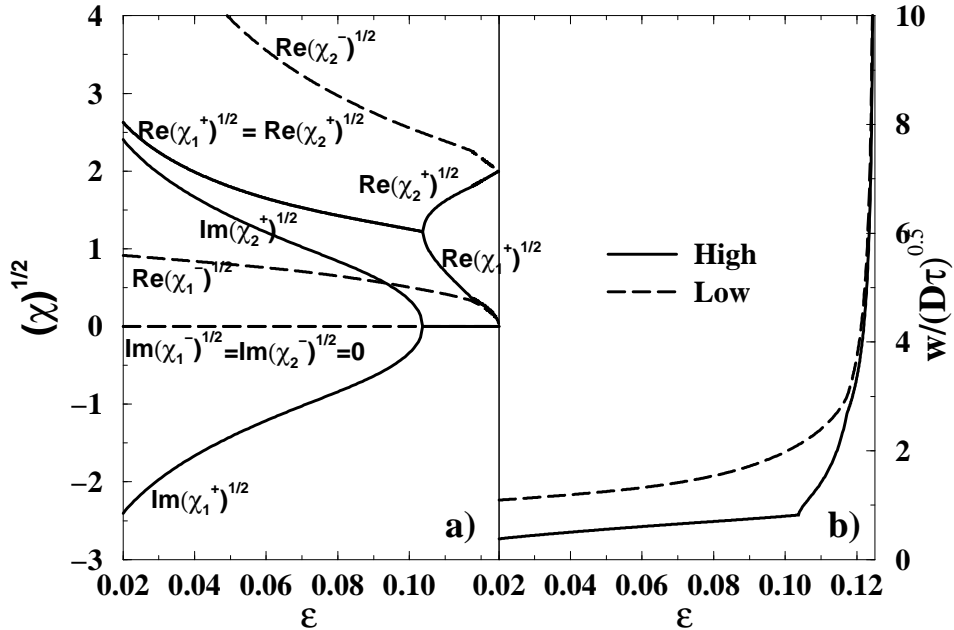}}}
\caption{(a) Eigenvalues of ${\bf M}^{\pm}(\sigma_{sel})$ the matrix of
  the linearised steady flow equations, for the high (+) and low (-)
  shear rate bands and (b) interface width on the high or low shear
  rate side of the interface, as a function of the retardation
  parameter $\epsilon$.}
\label{fig:evals}
\end{figure}

Asymptotically, the interface profile can be approximated by a
combination of exponentials, and the widths of the interface towards
the high shear rate band $w^+$ and towards the low shear rate band
$w^-$ are the characteristic decay lengths of the slowest of the
exponentials on the two sides of the interface, related to the
eigenvalues of ${\bf M}^{\pm}$:
\begin{equation}
w^{\pm} = \dfrac{\sqrt{{\cal D}\tau}}{\displaystyle\min_{i=1,2}\left\{
    \text{Re}\left(\sqrt{\chi^{\pm}_i (\hat{\sigma}^*,
      \epsilon)}\right)\right\}}
\end{equation}
where $i=1,2$ correspond to the two eigenvalues $\chi^{-}_{1,2}$ of
${\bf M}^{-}$ and to the two eigenvalues $\chi^{+}_{1,2}$ of ${\bf
  M}^{+}$ (see Fig.~\ref{fig:evals}a).  The dependence of the
interface width on the retardation parameter $\epsilon$ is shown in
Fig.~\ref{fig:evals}b. The interface is asymmetric (thicker in the low
shear rate band).

The non-vanishing imaginary parts of $\sqrt{\chi^{+}_{1,2}}$ imply
damped oscillations of the interface profile towards the high shear
rate band.  The wavelength of these oscillations is the inverse of
$|\text{Im}\sqrt{\chi^{+}_{i}}|$ and because
$|\text{Im}\sqrt{\chi^{+}_{i}}| <
\text{Min}(\text{Re}\sqrt{\chi^{+}_{i}})$ (Fig.~\ref{fig:evals}a)
the width of the interface is smaller compared to this wavelength the
overdamped oscillations are hardly noticeable in
Fig.~\ref{fig:cross}a showing the interface profile.

As seen in Fig.~\ref{fig:evals}a, a bifurcation occurs for low
polymer viscosity. For $\epsilon > 0.1$, the eigenvalues
$\chi_{1,2}^+$ are no longer complex conjugate ($\text{Re} \chi_1^+
\neq \text{Re} \chi_2^+$).  This bifurcation produces a discontinuity
of the first derivative of $w^+(\epsilon)$ at $\epsilon=0.1$. At the
same time the imaginary parts of $\chi_{1,2}^+$ vanish. This
particularity of the JS model was also noticed in \cite{malkus91}.  At
$\epsilon=1/8$, which is the limit of existence of banded solutions
(above $\epsilon=1/8$ the local constitutive curve is monotonic)
$\chi_1^-, \chi_1^+$ vanish, therefore both widths $w^{\pm}$ diverge,
Fig.~\ref{fig:evals}b.

The interface width scales like $\sqrt{{\cal D}\tau}$, which is
typically of the same order as the polymer chain size
\cite{eurorheo}.  The prefactor can be rather big for small polymer
viscosity (it diverges at the critical point $\epsilon = 1/8$ like
$(1/8 - \epsilon)^{-1/2}$) and it is geometry independent. Thus, the
width of a thin, steady interface should be the same in planar shear,
as in Poiseuille and Couette flows for any curvature.
Ref.~\cite{BritCall97} determined the width of the interface for
micelles experiments in cone and plate geometry, and reported an
increase in width of a factor 3 when the cone-plate angle changes from
$4^o$ to $7^o$.  Although we did not investigate this particular
geometry here, we expect that rather generally the width of a thin,
steady interface depends only on the constitutive model and does not
depend on the geometry.  The widths reported in \cite{BritCall97} are
much larger than the polymer chain size, therefore they are most
probably produced by a different mechanism. A possible explanation is
the dynamical broadening due to travelling surface waves in two-layer
flows \cite{renren93,khomami}.
\section{Conclusions} \label{sect:conclusions}
We have discussed the steady banded flow of the Johnson-Segalman
model, considering 1D shear rate profiles in Poiseuille, cylindrical
Couette and planar shear geometries, under controlled shear rate
conditions.  The introduction of a non-local diffusion term in the JS
local constitutive relation lifts the continuous degeneracy and the
history dependence of the steady flow solution for Poiseuille and
cylindrical Couette geometries, and generally for any geometry that
imposes a non-uniform total shear stress.  In order to analyse these
phenomena we used matched asymptotic techniques which are common in
the field of reaction-diffusion systems describing processes like
combustion or nerve conduction, but which, to our knowledge, are new
in the field of creeping flow of complex fluids.  The use of the JS
model in this paper should be considered only as an example of
application of these methods in order to investigate the consequences
of diffusion terms on banded flow.

We showed that in Poiseuille and cylindrical Couette geometries steady
banded flow has two bands, which seems to be the case in birefringence
experiments \cite{DCC97,decruppe98}.  NMR visualisation of more than
two bands as reported in Ref.~\cite{MairCall97} in the Couette
geometry could be due to slowly decaying (or pinned) transients. Note
that the speed of a moving interface scales as $D^{1/2}$.  Another
possibility, not taken into account here but that needs further
investigation, is that concentration effects may stabilise banded
flows with more than two bands.  Thus, a depletion layer at the inner
cylinder increases locally the selected stress and may equilibrate a
second interface closer to the inner cylinder than the first
equilibrium position which corresponds to higher concentration and
lower selected stress.

In the absence of diffusion steady banded flow can form any number of
bands.  The same arbitrariness of the number of bands was shown in
steady Poiseuille flow \cite{geovla98,fyrilas99}.  In numerical
simulations of Ref.~\cite{spenley96} no selection was noticed, except
for the one imposed by the mesh size.  Recently, we showed numerically
\cite{olmradlu99} that even in Couette flow, solving the flow in the
absence of diffusion is an ill-posed problem, unstable to noise in the
initial conditions.  The lack of robustness of the flow increases when
the curvature decreases, in the sense that larger and larger regions
inside the gap may split into multiple bands under perturbations of
the initial conditions.  Refs.~\cite{espanol96,yuan99} found an
unique, single interface solution even in the planar shear of the JS
model without a diffusion term, although this interface has a finite
width that suggests diffusion terms intrinsic to the finite element
scheme.  With diffusion the system of dynamical equations becomes
parabolic leading to compactness of the global attractors and to
stability with respect to initial conditions \cite{brunovsky94}.

In certain conditions in cylindrical Couette flow (low curvature and
high polymer viscosity), both normal and inverted band sequences are
allowed for the same shear rate. Nevertheless, the inverted band is
metastable and can be reached from rest only on the second half of the
plateau \cite{olmradlu99}.  Although this branch has not been observed
experimentally, its existence and linear stability is proven
rigorously and may have consequences on the kinetics of shear-banding
within a restricted domain of control parameters.

An interesting feature shown by experimental flow curves and
reproduced by our theoretical model is the presence of a selected
stress plateau, reminiscent of isotherms at liquid-gas or
decompositions of binary systems first order phase transitions.  In
thermodynamic transitions the plateau is fixed by a Maxwell type
construction, justified by the equality of the chemical potentials of
the coexisting phases. In the model we present here the equilibrium of
"phases" is a condition of existence of a stationary interface, and
relies on the balance between stress diffusion across the interface
and non-linear stress relaxation.  The retardation parameter plays the
r\^{o}le of temperature, and its critical value $\epsilon=1/8$ is
analogous to an equilibrium critical point. The extremities of the
plateau for varying $\epsilon$ describe a curve that is analogous to a
binodal for phase separation, the critical point being the top of the
binodal (Fig.~\ref{fig:localconst}a).  As in first order transitions
the width of the interface separating the two bands diverges at the
critical point.  Whether the studied phenomena have more than a formal
connection to thermodynamics is still an open question.

Apart from settling some basic questions concerning how the presence
of diffusion terms affects measurable flow curves, we hope that the
results presented here will stimulate more accurate and diversified
experiments, eventually searching for various unusual features such as
inverted band sequences and the special kinetic pathways involved.
This could be a test of the theoretical assumptions but could also
increase the operational range in view of possible applications.
\begin{ack}
  We thank T.C.B.~McLeish and C.-Y. D. Lu for useful discussions.
  This work is funded by EPSRC (GR/L70455).
\end{ack}
\appendix
\section{Stability of the flow with respect to
  changes in the position of the interface}
\label{appendix1}

Stability with respect to the position of the interface means that any
displacement $\delta \hat{r}^*$ of the interface leads to a moving
interface solution with velocity oriented towards the equilibrium
position $\hat{r}_{sel}$, {\sl i.e.\/}:
\begin{equation}
  c\left[\hat{\sigma}\left(\hat{r}_{sel}+\delta
      \hat{r}^*\right)\right] \delta \hat{r}^* =  
  \left.\frac{dc}{d\hat \sigma}\right|_{\hat{\sigma}_{sel}}
  \left.\frac{\partial \hat \sigma}{\partial 
    \hat{r}^*}\right|_{\hat V} (\delta \hat{r}^*)^2 < 0
  \label{eq:condition}
\end{equation} 

The partial derivative occurring in Eq.~(\ref{eq:condition}) can be obtained as
\begin{equation}
  \left.\frac{\partial \hat{\sigma}^* }{ \partial \hat{r}^* }\right|_{\hat V} =
  -\frac{\left.\frac{\partial \hat V}{\partial
        \hat{r}^*}\right|_{\hat{\sigma}^*}} 
  {\left.\frac{\partial \hat V}{\partial \hat{\sigma}^*}\right|_{\hat{r}^*}} 
  \label{eq:partdiff}
\end{equation}
Using Eqs.~(\ref{eq:partdiff}, \ref{eq:v-+poise}, \ref{eq:stresspoise})
for Poiseuille flow we obtain:
\begin{gather}
  \label{eq:dvdfp}
  \left.\frac{\partial \hat{V}}{\partial f}\right|_{\hat{\sigma}^*} = 
  \frac{\hat{\dot{\gamma}}^+(f) -\hat{V}}{f}  \\  
\label{eq:dsigmadr}
  \left.\frac{\partial \hat{\sigma}^*} {\partial \hat{r}^*}\right|_{\hat V} = 
  \frac{-\left.\frac{\partial \hat{V}}{\partial
      f}\right|_{\hat{\sigma}^*}\left.\frac{\partial f}{\partial 
      \hat{r}^*}\right|_{\hat{\sigma}^*}}{\left.\frac{\partial \hat{V}}{\partial
      \hat{\sigma}^*}\right|_{f}+\left.\frac{\partial \hat{V}}{\partial
      f}\right|_{\hat{\sigma}^*}\left.\frac{\partial f}{\partial
      \hat{\sigma}^*}\right|_{\hat{r}^*}}    = \frac{f}
  {1 -  P(f,\hat{\sigma}_{sel}) } \\
\intertext{where}
  P(f,\hat{\sigma}_{sel}) = 
  \frac{\hat{\sigma}_{sel} [ \hat{\dot{\gamma}}^+(\hat{\sigma}_{sel}) - 
    \hat{\dot{\gamma}}^-(\hat{\sigma}_{sel}) ]}
  {f[ \hat{\dot{\gamma}}^+(f) - \hat{V}(\hat{\sigma}_{sel},f)] }
\end{gather}

According to the tested conjecture, an increase of the total stress
will displace the interface towards the middle of the slit or pipe
($\frac{d c}{d \hat{\sigma}} < 0$), so using Eq.~(\ref{eq:dsigmadr})
the stability condition (\ref{eq:condition}) for plug flow in
Poiseuille geometry reads:
\begin{equation}
  P(f,\hat{\sigma}_{sel}) < 1.
\end{equation} 
Using the fact that $\hat{\dot{\gamma}}^-(\hat{\sigma})$ is an
increasing function of $\hat{\sigma}$ it follows from
Eq.~(\ref{eq:v-+poise}) that
$\hat{V}(\hat{\sigma}_{sel},\hat{\sigma}_{sel}) <
\hat{\dot{\gamma}}^-(\hat{\sigma}_{sel}) <
\hat{\dot{\gamma}}^+(\hat{\sigma}_{sel})$ and therefore
$P(\hat{\sigma}_{sel},\hat{\sigma}_{sel}) < 1$.
$P(f,\hat{\sigma}_{sel})$ is a monotonically decreasing function of
$f$, provided that $\hat{\dot{\gamma}}^+(\hat{\sigma})$ is increasing
with $\hat{\sigma}$, as shown by
\begin{equation}
\left.\frac{\partial P}{\partial f}
\right|_{\hat{\sigma}_{sel}} = - \frac{d \hat{\dot{\gamma}}^+(f)}{d f}
\left[\frac{P}{\hat{\dot{\gamma}}^+(f)-\hat{V}(\hat{\sigma}_{sel},f)}\right]
< 0.  
\end{equation}
Because the lowest allowed value of $f$ for banded flow is
$f=\hat{\sigma}_{sel}$, the stability criterion is always satisfied.

A similar stability analysis can be performed in the case of the
cylindrical Couette flow.  Using Eqs.~(\ref{eq:partdiff}),
(\ref{eq:v+-couette}), (\ref{eq:v-+couette}), (\ref{eq:stresscouette})
for Couette flow we obtain:
\begin{subequations}
\begin{gather}
  \label{eq:dv+-gamma}
  \left.\frac{\partial \hat{V}^{+-}}{\partial \Gamma}\right|_{\hat{\sigma}^*} = 
  \frac{\hat{\dot{\gamma}}^+(\Gamma) -  \hat{\dot{\gamma}}^-(\Gamma/(1+p)^2) }{2 \Gamma}  \\
  \label{eq:dv-+gamma}
  \left.\frac{\partial \hat{V}^{-+}}{\partial \Gamma}\right|_{\hat{\sigma}^*} = 
  \frac{ - \hat{\dot{\gamma}}^+(\Gamma /(1+p)^2) + \hat{\dot{\gamma}}^-(\Gamma) }{2 \Gamma}  \\
  \left.\frac{\partial \hat{\sigma}^*} {\partial
      \hat{r}^*}\right|_{\hat{V}^{+-}}  
  =  \frac{-4 \hat{\Gamma}^{-1/2} \hat{\sigma}_{sel}^{3/2}  }
  {1-C^{+-}(\hat{\Gamma},\hat{\sigma}_{sel},p) } \\
  \left.\frac{\partial \hat{\sigma}^*} {\partial
      \hat{r}^*}\right|_{\hat{V}^{-+}}  
  =  \frac{-4 \hat{\Gamma}^{-1/2} \hat{\sigma}_{sel}^{3/2}  }
  {1-C^{-+}(\hat{\Gamma},\hat{\sigma}_{sel},p) }, \\
\intertext{with}
  C^{+-}(\hat{\Gamma},\hat{\sigma}_{sel},p) = 
  \frac{ \hat{\dot{\gamma}}^+(\hat{\sigma}_{sel}) - 
    \hat{\dot{\gamma}}^-( \hat{\sigma}_{sel} ) }
  { \hat{\dot{\gamma}}^+(\hat{\Gamma}) - 
    \hat{\dot{\gamma}}^-(\hat{\Gamma}/(1+p)^2)  } \\
  C^{-+}(\hat{\Gamma},\hat{\sigma}_{sel},p) = 
  \frac{ \hat{\dot{\gamma}}^+(\hat{\sigma}_{sel}) - 
    \hat{\dot{\gamma}}^-( \hat{\sigma}_{sel} ) }
  {\hat{\dot{\gamma}}^+(\hat{\Gamma}/(1+p)^2) - 
    \hat{\dot{\gamma}}^-(\hat{\Gamma})  },
\end{gather}
\end{subequations}
where the sequence $+-$ refers to the high shear rate band near the
inner cylinder and the sequence $-+$ refers to the high shear rate
band near the outer cylinder.

The stability criteria for the two sequences are respectively:
\begin{subequations}
  \begin{align}
    C^{+-}(\hat{\Gamma},\hat{\sigma}_{sel},p) &< 1 \\
    C^{-+}(\hat{\Gamma},\hat{\sigma}_{sel},p) &> 1
  \end{align}
\end{subequations} 
Because $\hat{\Gamma}/(1+p)^2 < \hat{\sigma}_{sel} < \hat{\Gamma}$,
and provided that both $\hat{\dot{\gamma}}^-(\hat{\sigma})$, and
$\hat{\dot{\gamma}}^+(\hat{\sigma})$ are increasing functions of the
total stress $\hat{\sigma}$, the stability criteria are always
fulfilled for both sequences of bands.

To conclude, banded flow is stable with respect to displacements of
the interface in Poiseuille slit and pipe geometry.  In cylindrical
Couette flow, both sequences of bands are stable with respect to
displacements of the interface, provided that their existence is
allowed by further criteria discussed in Appendix~\ref{appendix2}.

\section{Existence conditions for band sequences in cylindrical
  Couette geometry} 
\label{appendix2}

The homogeneous low shear rate branch is obtained from
Eq.~(\ref{eq:v+-couette}) with $\hat{\sigma}^*=\hat{\Gamma}$
(interface at the inner wall $\hat{r}^*=1$ ), and according to
Eq.~(\ref{eq:homlow}) with the restriction $0 < \hat{\Gamma}/\hat{r}^2
< \sigma_{top}(\epsilon), \forall \hat{r} \in (1,1+p)$, which is
equivalent to $0 < \hat{\Gamma} < \hat{\sigma}_{top}(\epsilon)$.  The
high shear rate branch can be obtained from Eq.~(\ref{eq:v+-couette})
with $\hat{\sigma}^*=\hat{\Gamma}/(1+p)^2$ (interface at the outer
wall $\hat{r}^*=1+p$), and according to Eq.~(\ref{eq:homhigh}) with
the restriction $\hat{\Gamma}/\hat{r}^2 > \sigma_{top}(\epsilon),
\forall \hat{r} \in (1,1+p)$, i.e $\hat{\Gamma} >
(1+p)^2\hat{\sigma}_{bottom}(\epsilon)$.

Banded flow with the two possible band sequences is described by
Eqs.~(\ref{eq:v+-couette}, \ref{eq:v-+couette}) with
$\hat{\sigma}^*=\hat{\sigma}_{sel}$.  Eq.~(\ref{eq:banded}) imposes
the limiting values of the torque:
\begin{equation}
\hat{\sigma}_{sel} < \hat \Gamma = \hat{\sigma}_{sel} (\hat{r}^*_{sel})^2 
< \hat{\sigma}_{sel}(1+p)^2
\label{eq:lim_torque}
\end{equation}
The two limiting values correspond to the steady interface touching
the inner ($\hat{r}^*_{sel}=1$) and the outer ($\hat{r}^*_{sel}=1+p$)
cylinder, respectively.

If Eq.~(\ref{eq:banded}) is fulfilled, Eq.~(\ref{eq:banded+-}) is
automatically fulfiled because it becomes equivalent to
$\sigma_{bottom}(\epsilon) < \sigma_{sel}(\epsilon) <
\sigma_{top}(\epsilon)$.  Thus the sequence of bands $(+-)$ (high
shear rate band at the inner cylinder) exists for all torque values
such that:
\begin{equation}
  \hat \Gamma \in I_{\Gamma}^s :=
  (\hat{\sigma}_{sel},\hat{\sigma}_{sel}(1+p)^2). 
\end{equation}
If Eq.~(\ref{eq:banded}) is fulfilled,
the condition~{\ref{eq:banded-+}} becomes equivalent to: 
\begin{equation}
  \hat \Gamma \in I_{\Gamma} :=
  ((1+p)^2\hat{\sigma}_{bottom},\hat{\sigma}_{top}),
\end{equation}
hence the $(-+)$ sequence of bands (high shear rate band at the outer
cylinder) exists for torques $\hat \Gamma \in I_{-+} = I_{\Gamma}^s
\cap I_{\Gamma}$.  The interval $I_{-+} $ can be either empty, or
strictly included in $I_{\Gamma}^s$, or coincide with $I_{\Gamma}^s$.
There are several regions in the plane of parameters $(\epsilon, p)$,
corresponding to these situations.

\renewcommand{\theenumi}{\Roman{enumi}}
\begin{enumerate}
\item $\hat{\sigma}_{bottom}(\epsilon) (1+p)^2 < \hat{\sigma}_{sel}$,
  $\hat{\sigma}_{top}(\epsilon) > \hat{\sigma}_{sel}(1+p)^2$,
  $\hat{\sigma}_{bottom}(\epsilon)(1+p)^2 <
  \hat{\sigma}_{top}(\epsilon)$.  In this region both band sequences
  are allowed for torques in the full interval $I_{\Gamma}^s$ because
  $I_{\Gamma}^s \subset I_{\Gamma}$ and therefore $I_{-+}=I_{\Gamma}^s$.
\item $\hat{\sigma}_{bottom}(\epsilon) (1+p)^2 > \hat{\sigma}_{sel}$,
  $\hat{\sigma}_{top}(\epsilon) > \hat{\sigma}_{sel}(1+p)^2$,
  $\hat{\sigma}_{bottom}(\epsilon)(1+p)^2 <
  \hat{\sigma}_{top}(\epsilon)$.  In this region the sequence $(+-)$
  is allowed for torques within $I_{\Gamma}^s$, while the sequence
  $(-+)$ is allowed in the interval
  $I_{-+}=(\hat{\sigma}_{bottom}(\epsilon)(1+p)^2,\hat{\sigma}_{sel}(\epsilon)
  (1+p)^2)$ obtained by truncating $I_{\Gamma}^s$ at its low stress limit.
\item $\hat{\sigma}_{bottom}(\epsilon) (1+p)^2 < \hat{\sigma}_{sel}$,
  $\hat{\sigma}_{top}(\epsilon) < \hat{\sigma}_{sel}(1+p)^2$,
  $\hat{\sigma}_{bottom}(\epsilon)(1+p)^2 <
  \hat{\sigma}_{top}(\epsilon)$.  In this region the sequence $(+-)$ is
  allowed for torques within $I_{\Gamma}^s$, while the sequence $(-+)$ is
  allowed in the interval
  $I_{-+}=(\hat{\sigma}_{sel}(\epsilon),\hat{\sigma}_{top}(\epsilon))$
  obtained by truncating $I_{\Gamma}^s$ at its high stress limit.
\item $\hat{\sigma}_{bottom}(\epsilon) (1+p)^2 > \hat{\sigma}_{sel}$,
  $\hat{\sigma}_{top}(\epsilon) < \hat{\sigma}_{sel}(1+p)^2$,
  $\hat{\sigma}_{bottom}(\epsilon) (1+p)^2 < \hat{\sigma}_{top}(\epsilon)$.
  In this region the sequence $(+-)$ is allowed for torques within
  $I_{\Gamma}^s$, while the sequence $(-+)$ is allowed in the interval
  $I_{-+}=(\hat{\sigma}_{bottom}(\epsilon)(1+p)^2,\hat{\sigma}_{top}(\epsilon))$
  is obtained by truncating $I_{\Gamma}^s$ at both limits.
\item $\hat{\sigma}_{bottom}(\epsilon) (1+p)^2 > \hat{\sigma}_{sel}$,
  $\hat{\sigma}_{top}(\epsilon) < \hat{\sigma}_{sel}(1+p)^2$,
  $\hat{\sigma}_{bottom}(\epsilon) (1+p)^2 >
  \hat{\sigma}_{top}(\epsilon)$.  In this region the sequence $(+-)$ is
  allowed for torques within $I_{\Gamma}^s$, while the sequence $(-+)$ is
  forbidden for all values of the torque, because $I_{-+}=\emptyset$.
\end{enumerate}
\renewcommand{\theenumi}{\arabic{enumi}}

These five regions are shown in Fig.~\ref{fig:domains}.  Region III is
very narrow, and practically represents the frontier between the
Regions I and IV.  In general, for given $\Gamma$ and $\epsilon$ the
inverted band sequence is forbidden for sufficiently highly curved
geometries (large enough $p$).  A less formal approach to obtain the
same results is shown in Fig.~\ref{fig:sigma_profiles}.
Fig.~\ref{fig:sigma_profiles}c corresponds to the case of large
curvature (region IV), in which the value of the stress does not allow
for the existence of the low shear rate band at the outer cylinder
forbidding thus the sequence (-+), because $\Gamma/(1+p)^2 <
\sigma_{bottom}$.  For small curvature
(Fig.~\ref{fig:sigma_profiles}b) the values of the stress at the inner
and outer cylinders allow the existence of both high and low shear
rate bands, thus both sequences of bands can exist.  The possibility
of an inverted band sequence was also found in Ref.~\cite{goveas99}
for a two fluid model of shear banding in polymer blends.


\begin{thebibliography}{10}
\bibitem{comment} Examples (from biology and solid-state physics) of
  non-potential reaction-diffusion systems for which the dominance
  principle does not apply are described in Ref.~\cite{bode97}.  In
  these special systems propagation of the interface is possible in
  both directions, depending on the shape of the interface profile.
\end{thebibliography}

\end{document}